\documentclass[twocolumn,preprintnumbers,amsmath,amssymb]{revtex4}
\usepackage{epsfig}
\usepackage{dcolumn}
\usepackage{bm}

\newcommand{\LB}[1]{\label{#1}}

\begin{document}

\title{Many Attractors, Long Chaotic
Transients, and Failure in Small-World Networks of Excitable Neurons} 

\author{Hermann Riecke}
\affiliation{Department of Engineering Sciences and Applied Mathematics,
Northwestern University, Evanston, IL 60208, USA}
\author{Alex Roxin}
\affiliation{Computational Neuroscience,
Departament de Tecnologia,
Universitat Pompeu Fabra,
08003 Barcelona, Spain}
\author{Santiago Madruga}
\affiliation{Max-Planck-Institute for Physics of Complex Systems, D-01187
Dresden, Germany}
\author{Sara A. Solla}
\affiliation{ Department of Physiology and Department of Physics and
Astronomy, Northwestern University, Evanston, IL 60208, USA}

\date{\today} \maketitle 

\section*{Abstract} We study the dynamical states that emerge in a
small-world network of recurrently  coupled excitable neurons through
both numerical and analytical methods.  These dynamics  depend in
large part on the fraction of long-range connections or `short-cuts'
and the delay in the neuronal interactions.  Persistent activity
arises for a small fraction of `short-cuts',  while a transition to
failure occurs at a critical value of the `short-cut' density.  The
persistent activity consists  of multi-stable periodic attractors, the
number of which is at least on the order of the number of neurons in
the network. For long enough delays, network activity at high
`short-cut' densities is shown to exhibit exceedingly long chaotic
transients  whose failure-times averaged over many network
configurations follow a stretched exponential.  We show how this 
functional form arises in the ensemble-averaged activity if each
network realization  has a characteristic failure-time which is
exponentially distributed.

\bigskip

\textbf{Many systems in nature can be described as a network of
interconnected nodes.  A growing list of examples, from social and
ecological webs to the neural anatomy of simple organisms, have been
shown to exhibit  complex topological features in stark contrast to
ordered lattices or purely random networks. It is futhermore becoming
increasingly clear that  the details of the network architecture can
crucially influence the  emergent dynamics in the system.  An
understanding of the dynamics on complex  networks requires
investigation into the interplay between the instrinsic dynamics of
the elements at the nodes and  the topology of the network in which
they are embedded.  Here we study the collective behavior of excitable
model neurons  in a network with Small-world topology.  The
Small-world network is locally  ordered but includes a certain number
of randomly placed, potentially long-range connections  which may act
as short cuts.  Such a topology bears a schematic resemblence to that
found  in the cerebral cortex, in which neurons are most strongly
coupled to nearby  cells, but are known to make long-range projections
to cells millimeters away.  We show that in the regime of low
`short-cut'  density, persistent activity arises in the form of
propagating excitable  waves which annihilate upon meeting and are
spawned anew via re-injection  by long-range connections.  A critical
value of the `short-cut' density  at which network activity fails can
be found by matching the longest  distance on the network, a
topological measure, to an intrinsic recovery  time of the individual
cells.  We furthermore show how activity once again  emerges at high
`short-cut' densities if the delay in the neuronal interactions  is
sufficiently long.  The activity in this regime exhibits long, chaotic
transients composed of noisy, large-amplitude population bursts.  In
this  regime we investigate the interplay of the delay and the network
topology numerically  and provide a simple argument which provides a
qualitative description  of the distribution of failure times of the
chaotic network activity.}

\section{Introduction}
\LB{s:intro}

It has been widely recognized that the connectivity of a network of
active elements has a profound impact on its function. Substantial
effort has therefore been devoted to the characterization of the
network connectivity \cite{AlBa02,Ne03} leading to the identification
of various measures that are significant in determining the properties
of the system. Particularly relevant among them are the average and
maximal length of paths connecting arbitrary nodes in the network and
the distribution for the number of links (degree) that emanate from
the nodes.

The dynamics of networked elements has been studied in particular
detail for coupled oscillators, addressing the influence of the
network topology on their ability to synchronize. The existence of
long-range connections between oscillators, which reduces the
effective size of the network, has been found to enhance
synchronizability substantially \cite{BaPe02}. At the same time, the
heterogeneity of the degree distribution of many such networks limits
the ability of the oscillators to synchronize, requiring a balance
between the two aspects \cite{NiMo03}. 

Excitable elements constitute a second important class of dynamical
systems. In locally coupled networks they give rise to traveling waves
(e.g. \cite{Br00d,Br01,GoEr01}). If these waves annihilate upon mutual
collisions, which is typically the case, persistent activity  usually
requires an external drive or spontaneous excitation by noise. In
models for neural systems driven by noise \cite{TrSc99,LeRi00,LeRi01}
networks with non-local connections between their elements have been
shown to possess the tendency to exhibit relatively ordered
oscillations in the population (mean) activity. The spatial structure
of such noise-induced waves becomes less coherent with an increase in
the non-local coupling \cite{Pe05a}. Combining local connectivity with
a small number of non-local connections allows a localized
time-periodic external input to entrain the whole system much faster
than in a purely local network and at the same time the oscillations
are much more coherent than in a truly random network
\cite{LaHu00,LaCo01}. The dynamics depend very strongly on the
duration of the refractory period relative to the propagation time
scales and the range of the coupling. Thus, for very short refractory
periods the small loops that are sufficiently likely in many types of
random networks allow persistent activity even in the absence of noise
as long as the initial conditions are asymmetric enough
\cite{CaLa06}.  In an epidemic model the introduction of non-local
connections was found to induce a transition to a state with coherent
population oscillations \cite{KuAb01}.

In most studies the connections between the elements have been assumed to
be bi-directional, which is a very natural assumption in an epidemic
context \cite{KuAb01} and can be for neural systems if the neurons are
coupled via gap junctions \cite{LeRi00}. In the absence of noise
persistent activity is then only obtained with initial conditions that
suitably break the reflection symmetry. If in an initially quiescent
network individual neurons are excited by an external perturbation the
pair-wise excited symmetrical waves are, however, all eventually
annnihilated in collisions.  

In neural systems the coupling between neurons is predominantly not
bi-directional; instead most connections are via chemical synapses in
which usually the information is only transmitted from the
pre-synaptic axon to the post-synaptic dendrite. For these situations
it is appropriate to consider directed networks with uni-directional
connections. In previous work \cite{RoRi04} we have investigated
networks in which the local connections are bi-directional assuming
that due to the close phyiscal proximity the probability for
reciprocal axo-dendritic connections is quite high, but the non-local
connections (`short-cuts') are uni-directional. Considering this
directed modification of the by now classical small-world network
\cite{WaSt98}, we found that even a few such uni-directional short-cuts
allow for persistent activity even for generic localized excitations. As
the density of short-cuts is increased, however, more and more network
configurations support only a brief population burst after which the
activity dies out. For slow propagation speeds of the waves the
failure of network activity was found to be delayed and could occur
after many cycles of chaotic population bursts. The simplicity of this
model allowed an analytical result for the failure transition and
detailed numerical analysis. 

The results for the minimal system \cite{RoRi04} provided important
insight into the phenomena found in simlations of more elaborate
models that were motivated by concrete biological situations. In
\cite{NeCl04} the connection between the topology of a neural network
and its tendency towards epileptic seizures has been studied and
related to the degree of recurrent connectivity of different parts of
hippocampus. The origin of bursting behavior was addressed in
\cite{ShTs06}. Quite commonly bursting behavior arises when fast
spiking behavior drives a slow process that in turn can shut off the
spiking. Typically the slow process is associated with some slow
kinetics. In \cite{ShTs06} the authors show that no such slow kinetics
are needed in networks with small-world topology. Both the seizing
and the bursting activity can be interpeted quite well based on an
understanding of the failure transition in the basic model of
\cite{RoRi04}. The rapid spread of activity and the persistent
oscillations have also been observed in recent experiments on the
Belousov-Zhabotinsky reaction in which the uni-directional short-cuts
were implemented exploiting the photo-sensitivity of the reaction
\cite{TiCu05,StTi06}.

Here we build on our previous results and present a detailed
characterization of the persistent states and the dependence of their
properties on the prevalence of short-cuts as well as of the long
chaotic transients for which we provide an understanding of the
stretched exponential behavior in their failure times. In Section
\ref{s:model} we define the model, which consists of 
`Integrate-and-Fire' neurons that are coupled via excitatory pulses in
a small-world topology with uni-directional short-cuts. In Section
\ref{s:ordered} we discuss in detail the persistent states and the
cross-over from persistent activity to failure for the case of rapidly
propagating waves. In Section \ref{s:disorder} we provide a detailed
analysis of the exceedingly long chaotic transients in the regime of
slower propagation. In the concluding Section \ref{s:concl} we discuss
our results in light of other work on neural networks with small-world
topology \cite{NeCl04,ShTs06}.

\section{Neuron Model and Network Topology}
\LB{s:model}

We consider a one-dimensional network of $N$ identical
integrate-and-fire neurons. Their membrane voltage is described by

\begin{eqnarray}
\tau\frac{dV_{i}}{dt}&=&-V_{i}+
R I_{ext}+g_{syn}\sum_{j=1}^{N}w_{ij}\delta(t-t^{j}-\tau_{D}),
\LB{e:IF}
\end{eqnarray}
with the reset condition
\begin{eqnarray}
V(t^{+})&=&V_{res}\hspace{0.5in}
\textrm{whenever}\hspace{0.5in}  V(t^{-})=V_{th}, \LB{reset}
\end{eqnarray}

Here $\tau$ is the membrane time constant,  $g_{syn}$ is the synaptic
strength measuring the change in voltage due to one incoming action
potential, $w_{ij}=0$ or $1$ indicates the absence or presence
respectively of a synaptic connection from neuron $j$ to neuron $i$,
$t^{j}$ is the time at which neuron $j$ fires an action potential,
$I_{ext}$ is an external current and $R$ the membrane resistance. The
effective delay $\tau_{D}$ in the neuronal interaction includes the
time for the signal to propagate along the axon as well as the time
needed for initiating the triggered action potential. If the latter
time dominates over the axonal delay the dependence of the delay on
the distance between the neurons involved can be neglected.
Post-synaptic currents due to synaptic activation are considered
instantaneous and are therefore modeled as Dirac-delta functions. 
Eq.(\ref{reset}) reflects the fact that whenever the voltage of neuron
$i$ reaches the threshold value $V_{th}$, a spike is emitted and the
voltage is reset to the value $V_{res}$. We write Eq.{\ref{e:IF} in
terms of dimensionless quantities by setting without loss of
generality $V_{res}=0$, rescaling $V$ and $g_{syn}$ by $V_{res}$ and
time by $\tau$ \footnote{In \cite{RoRi04} we had taken $\tau=10$.},
and by replacing $I_{ext}$ by the steady-state voltage $V_\infty$,
to which it corresponds in the absence of synaptic input. We focus on
the case of excitable rather than spontaneously oscillating neurons,
i.e. $V_\infty<1$, and consider only initial conditions in which at
most a few neurons are triggered to spike with all other neurons
quiescent. In the absence of noise this implies that neurons can only
fire at times that are multiples of $\tau_D$ and the model can be
solved exactly between these times. In the numerical computations the
time step is therefore taken to be $\Delta t=\tau_D$.

In cortex neurons receive often not only local input from neurons
near-by but also input from some distant neurons through long-range
projections. We mimic such a heterogeneous connectivity with an
extremely simplified architecture for the network in which each neuron
is connected to $2k$ neighbors, i.e. $w_{ij}=1$ for $j\le i\pm k$ and
$j\ne i$ and an {\em additional} $pN$  neurons that are chosen randomly. The
parameter $p$ thus gives the density of random uni-directional
connections as a fraction of the total number of neurons $N$. 

The dynamics arising in Eqs(\ref{e:IF},\ref{reset}) for purely
local connectivity ($p=0$) depend,  in the absence of noise, on the
interplay between the strength of the synapses $g_{syn}$, the number
of neighbors $2k$, and the delay $\tau_D$. This is most easily
illustrated for nearest-neighbor connectivity ($k=1$). In this case,
if $g_{syn}+V_\infty<1$ the pre-synaptic input is insufficient to cause
a spike and no network activity will occur. For stronger input, i.e.
for $g_{syn}+V_\infty>1$, a wave of speed $1/\tau_{D}$ is generated.
After the spike the voltage is reset to $V=0$ and the neuron is only
ready to fire again after it has recovered to the extent that an input
of magnitude $g_{syn}$ is sufficient to trigger another spike. This
recovery time is given by 

\begin{equation}
T_{R}=\ln{\Bigg(\frac{V_\infty}{V_\infty+g_{syn}-1}\Bigg)}  \LB{e:Tr}
\end{equation}

Note, that $T_R$ is not an intrinsic refractory period of
the neuron since for sufficiently strong input $g_{syn}$ this model
neuron can fire arbitrarily fast. 

Due to the bidirectionality of the local connections the firing of
each neuron not only triggers a spike in the neuron ahead of it in the
wave of excitation, but gives also an input to the neuron behind it.
Thus, the latter neuron receives already a first input at a time 
$2\tau_D$ after its firing. If $T_R> 2\tau_D$ this input is not
sufficient to trigger a spike and the activity propagates away from
the site of initiation as each neuron contributes exactly one spike,
see Fig.\ref{f:singlewave}a.  If, however, $T_R\le 2\tau_D$  then the
wave front entrains all the neurons in its wake, eventually leading to
synchronized activity of the whole network. Unless autapses with
$w_{ii}\ne 0$ are included the network breaks up into two synchronous
groups of neurons that fire out of phase with one another, see
Fig.\ref{f:singlewave}b. In the general case, a propagating wave can
be sustained if
$g_{syn}e^{\tau_{D}}\sum_{n=1}^{k}e^{-n\tau_{D}}+V_\infty>1$.  That is,
a neuron receives inputs from $k$ neighbors as the wave approaches,
with an input at a distance $n$ discounted by $e^{(1-n)\tau_{D}}$. 
For simplicity we will focus on the case of  nearest-neighbor coupling
($k=1$) , and will consider the regime in which waves of excitation
propagate, but do not entrain activity in their wake. This choice
constrains the allowable values of $V_\infty$ and $g_{syn}$. 
Specifically, we take $V_\infty=0.85$ and $g_{syn}=0.2$ unless
otherwise noted. In this regime the collision of two waves leads to
their mutual annihilation and after having fired in a propagating wave
a neuron can be re-excited by a single input of size $g_{syn}$ after a
time 

\begin{equation}
T_{R}^{(1)}=\ln{\Bigg(\frac{V_\infty-g_{syn}e^{2\tau_D}}{V_\infty+g_{syn}-1}\Bigg)}.
\LB{e:Tr1}
\end{equation}
This time includes the input after a delay of $2\,\tau_D$ from the neuron
further ahead in the wave.

\begin{figure}
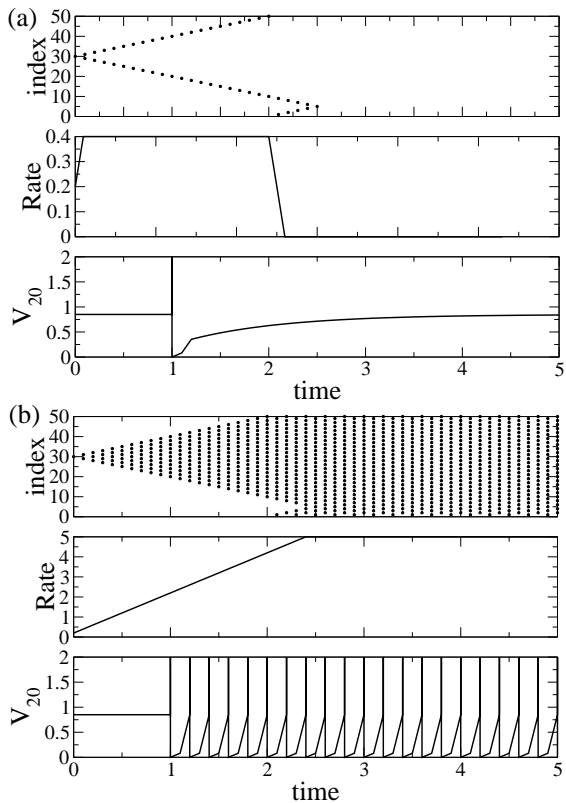

\begin{center}
\includegraphics[scale=0.3]{singlewave.eps}\\
\includegraphics[scale=0.3]{singlewave2.eps}
\end{center}
\caption[]{Dynamics in the flocal network of 50 neurons with
nearest-neighbor coupling.  Parameters are $V_\infty=0.85$ and
$\tau_{D}=0.1$.  (a) $g_{syn}=0.2$: An example of a propagating wave.
The wave fronts meet and annihilate at time 2.5.  Top: Raster plot
showing the times at which the  neurons fire an action potential.
Middle: The averaged firing rate of the network.  Bottom: The voltage
of neuron number 20.   (b) $g_{syn}=1.0$: The wave entrains all the
neurons in its path.  The final state of the network is one in which
two  synchronous groups  of neurons fire out of phase with one
another.  Note: The maximum possible firing rate is $1/\tau_{D}=10$.} 
\LB{f:singlewave}
\end{figure}

Eqs(\ref{e:IF},\ref{reset}) with the above-mentioned assumptions are a
 minimal model for the generation and propagation of waves in 
 cortical-like tissue.  As we shall see, the addition of random,
 long-range connections qualitatively  alters the dynamics of the
 network, allowing for a rich variety of spatio-temporal patterns.

\section{The ordered regime: Attractors and Failure}
\LB{s:ordered}

Once the input current $V_\infty$ and synaptic strength $g_{syn}$ have
been fixed, the dynamics arising in Eqns(\ref{e:IF}) and (\ref{reset})
are determined by the remaining two parameters: the fraction of
randomly placed connections $p$ and the delay in the  neuronal
interaction $\tau_{D}$.

\begin{figure}
\begin{center}
\includegraphics[scale=0.7]{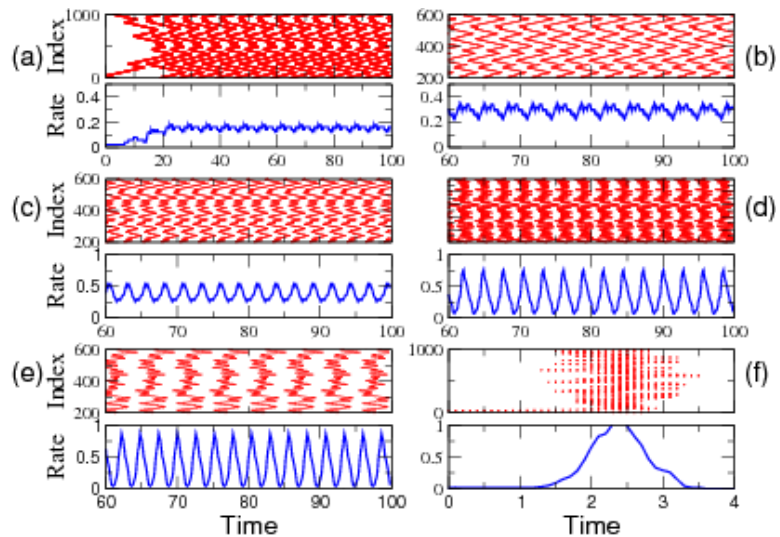}
\end{center}
\caption[]{Examples of the network dynamics as a function of the 
density of short-cuts $p$.  Here $p=0.01,0.05,0.1,0.15,0.2,0.25$ for
(a) through (f) respectively.  $\tau_{D}=0.1$.} \LB{f:increasep}
\end{figure}

The dynamics for non-vanishing $p$ differ qualitatively from those
without any long-range connections. The presence of long-range
connections (`short-cuts') allows the waves of excitation to be
re-injected into portions of the network which have already been
excited.  This process of re-injection may lead to persistent network
activity as shown in Fig.\ref{f:increasep}. As the waves spread
outward from the initial site of stimulation they encounter long-range
connections and are injected elsewhere in the network. Wavefronts that
meet  annihilate.  After some time, the activity settles into a stable
pattern in which the rates of generation and annihilation of the waves
balance. Averaged over time and across configurations, the firing rate
of these persistent states increases rapidly with increasing number of
short-cuts and saturates around $p \sim 0.1$ (Fig.\ref{f:firingrate}).
This saturation is a consequence of the neuron's finite recovery
period $T_R^{(1)}$ given in Eq.\ref{e:Tr1}. The bars in
Fig.\ref{f:firingrate} give the standard deviation of the firing rate
across configurations for $\tau_D=0.1$. Clearly in some configurations
the firing rate comes very close to its maximal value. As  the system
size is increased this saturation level is reached already for smaller
values of $p$ (Fig.\ref{f:firingrate}b). 

The firing rate is essentially the inverse of the time between
successive waves passing through a given point. Thus, one may expect
that decreasing the wave speed by increasing $\tau_D$ would reduce the
mean firing rate. This is, however, not the case; instead the firing
rate is quite insensitive to the wave speed as shown in
Fig.\ref{f:firingrate}b. The reason for this is apparent in
Fig.\ref{f:morewaves}a,b. It shows two simulations for the same
network configuration but with different delay times $\tau_D$. For
larger delay $\tau_D$  additional waves are excited by the short-cuts
as can be seen, for instance, at $t=380 \tau_D$ where a new wave can
be spawned at neuron 1071 for $\tau_D=0.1$ but not for $\tau_D=0.05$
(marked by circles). This increases the firing rate, and in the case
shown it even over-compensates the reduced wave speed, as can be seen
by the higher density of waves for $\tau_D=0.1$ at larger times.

\begin{figure}
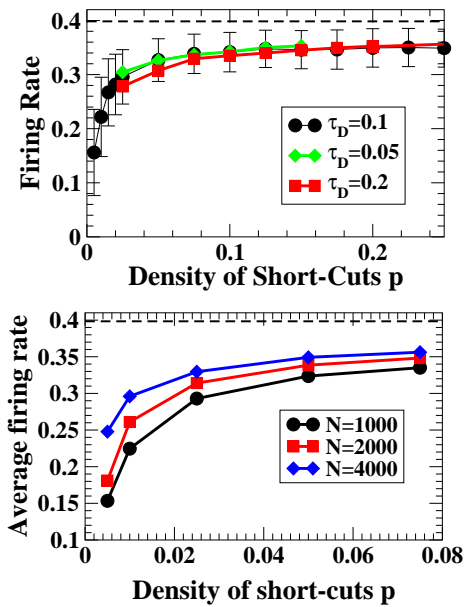

\begin{center}
\includegraphics[scale=0.4]{firingrate.ptaud.eps}\\
\vspace*{.2cm}
\includegraphics[scale=0.4]{firing.pN.eps}
\end{center}
\caption[]{The firing rate is close to $1/T_R^{(1)}$.
The firing rate averaged over 1000 configurations (5,000 for $p > 0.2$) 
as a function of the density of shortcuts $p$ for different values of the 
delay $\tau_D$ ($N=1000$) (a) and for different values of the system
size $N$ ($\tau_D=0.1$) (b). 
The theoretical maximal valueof $1/T_{R}^{(1)}$ is indicated by the dashed 
line. The bars indicate twice the standard deviation across configurations.} 
\LB{f:firingrate}
\end{figure}   

\begin{figure}
\begin{center}
\includegraphics[scale=0.5]{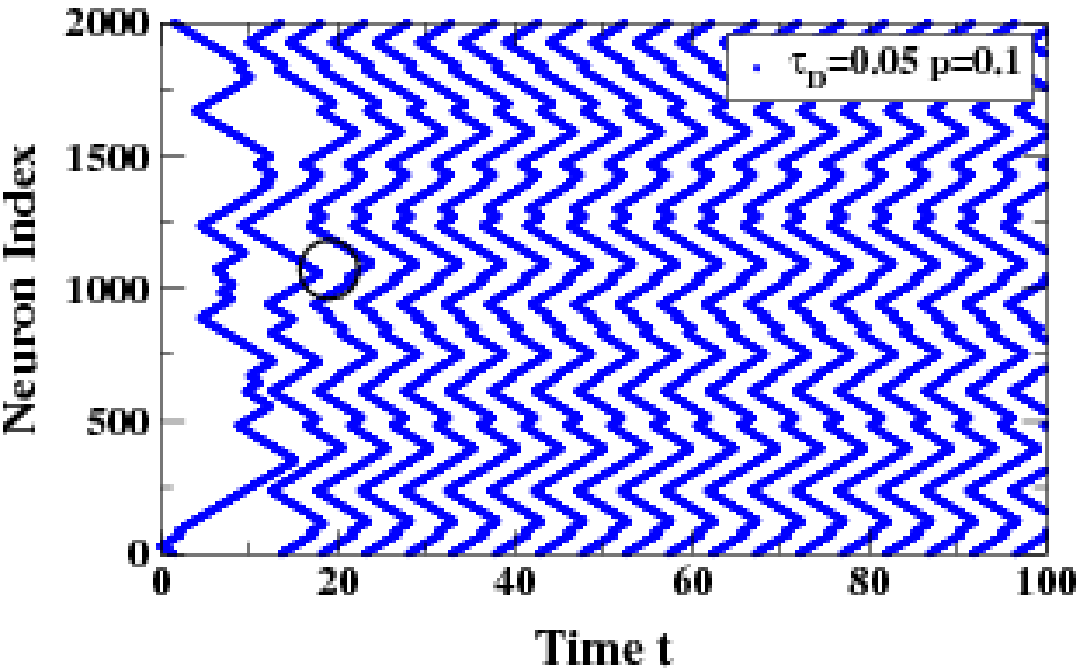}
\includegraphics[scale=0.5]{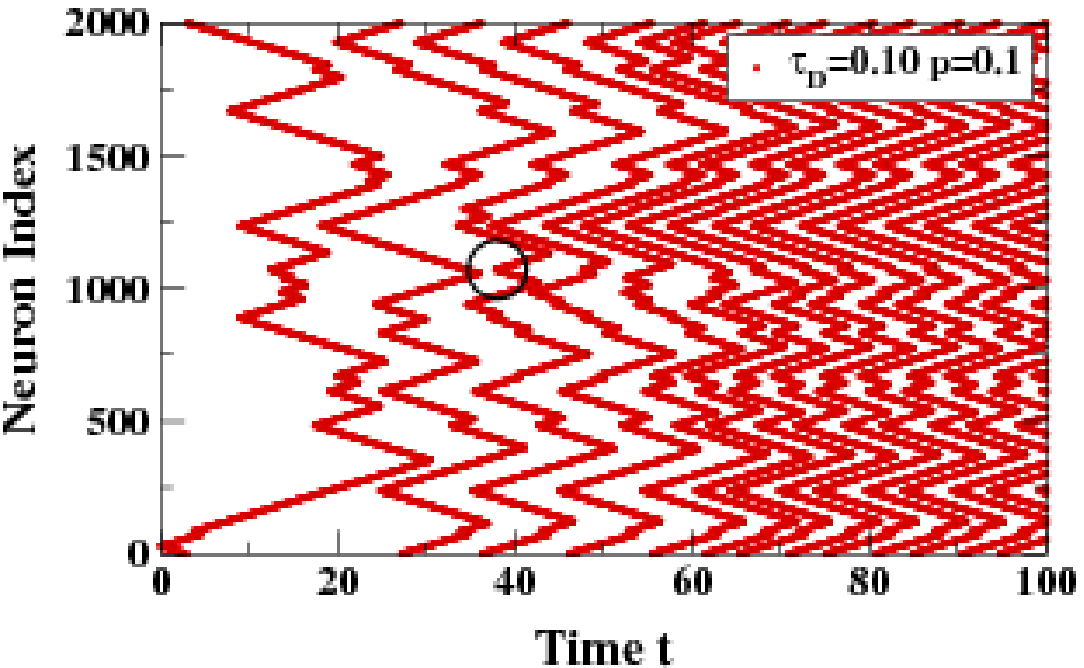}
\end{center}
\caption{For longer delay in an identical network, more waves can travel through the system.
Raster plots for $\tau_D=0.05$ and $\tau_D=0.1$ with $p=0.1$ and
$N=2000$. Circles mark where a new wave is spawned for $\tau_D=0.1$
but not for $\tau_D=0.05$.}
\LB{f:morewaves}
\end{figure}

For low $p$ only a few short-cuts are present and many pathways
leading to persistent activity consist only of large closed loops,
implying low firing rates. As $p$ increases the typical loop sizes
decrease and network configurations with large minimal loops and
correspondingly low firing rates become increasingly unlikely. This is
seen in the distribution function for the firing rate shown in
Fig.\ref{f:distfiringrate}. As the distribution function is shifted
towards larger firing rates with increasing $p$ it narrows
substantially reflecting the maximal firing rate set by the recovery
time $T_R^{(1)}$, which is marked by a dashed line in
Fig.\ref{f:distfiringrate}. 

\begin{figure}[h] 
\begin{center}
\includegraphics[scale=0.4]{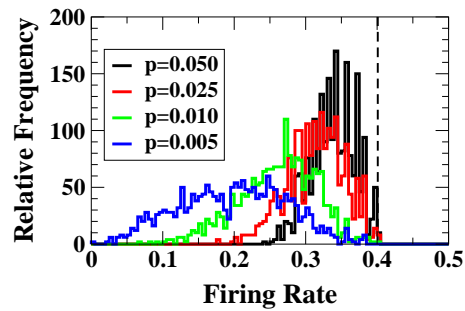} 
\end{center}
\caption{Distribution of firing rates across configurations for
various values of $p$.  As $p$ increases the distributions narrow in
width and approach the value $1/T_{R}^{(1)}$ (dashed line). $N=2,000$,
$\tau_D=0.1$}
\LB{f:distfiringrate} 
\end{figure}    

The activity into which the network eventually settles is periodic in
time and with increasing number of short-cuts the amplitude of the
oscillations increases. This is shown quantitatively in
Fig.\ref{f:osci}, which characterizes the magnitude of the
oscillations by the standard deviation of the firing rate averaged
over a large number of  persistent configurations. As is apparent from
Fig.\ref{f:increasep}, the time for these oscillations
to establish themselves from the excitation of a single neuron 
decreases with increasing $p$. This corresponds to the main result of
\cite{LaHu00,LaCo01} where it was found that in the small-world regime
a network of Hodgkin-Huxley neurons and of FitzHugh-Nagumo neurons is
entrained much more quickly by a cluster of driven, oscillating neurons.

\begin{figure}
\begin{center}
\includegraphics[scale=0.4]{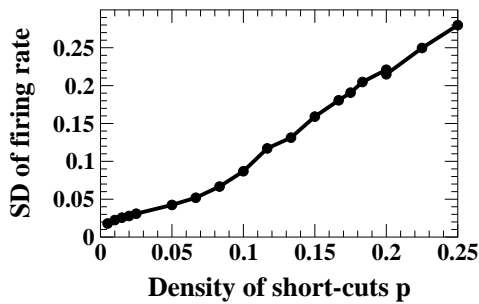}
\end{center}
\caption[]{Oscillation amplitude increases with the number of
short-cuts. Dependence of the standard deviation of the firing rate
(for persisting activities) averaged over up to 5,000 configurations
for $\tau_{D}=0.1$. $N=1000$.}\LB{f:osci}
\end{figure}

For larger values of $p$ the activity patterns can be quite
complicated.  While in this regime all neurons get excited during an
oscillation cycle, not all neurons and connections between them are
necessary for the persistence of the activity. This is illustrated in
Fig.\ref{f:pathway}. The full raster plot is depicted in the top panel
(black dots). Superimposed on it are those neurons that are essential
for carrying on the activity; they constitute the pathway of this
activity \cite{Ro03}. It is is determined as follows. Once the
simulation has been run, all the neurons that fire at the final time
are labeled. For each of the labeled neurons, a search is carried out
for presynaptic active neurons, i.e. neurons that fired one delay
$\tau_D$ ago and triggered the currently firing neurons. These
`causal' presynaptic neurons are labeled in turn. This process is
carried on backwards in time until $t=0$. The pattern of labeled
neurons quickly converges to a small subset, see middle panel of
Fig.\ref{f:pathway}. Only the neurons in this much simpler subset
contribute to the persistence of the pattern, i.e. the remaining
neurons could be cut out of the network without destroying the
persistent activity. In Fig.\ref{f:pathway}, the pathway of activity
(shown blown up in the bottom panel) is periodic in time with a period
that is only slightly longer than the recover time $T_{R}^{(1)}$,
indicated by the dashed lines. Such pathways have also been identified
in an experimental study of the photo-sensitive, excitable
Belousov-Zhabotinsky reaction in which short-cuts were introduced
through local optical excitation \cite{StTi06}.

\begin{figure}
\begin{center}
\includegraphics[scale=0.5]{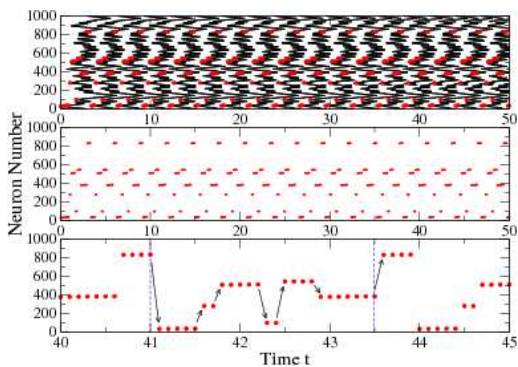}
\end{center}
\caption[]{An example of a 'pathway' of excitation.  
The pathway responsible for persistence is shown in red and is 
longer than the recovery time $T_{R}^{(1)}$, indicated by the dashed lines in the lower panel.  
$\tau_{D}=0.1$ and $p=0.1$.  See text for details.} \LB{f:pathway}
\end{figure} 

A striking feature of this regime is the fact that for a given network
configuration there is an extraordinarily large number of different
attractors. To assess the degree of multiplicity of attractors we
focus on the regime with low short-cut density where all solutions are
periodic and characterize each solution by its period, its mean firing
rate, and the standard deviation of the firing rate. Clearly,
distinguishing different solutions by only these three measures may
underestimate the total number of attractors. Fig.\ref{f:multatt}
illustrates the attractor multiplicity for a single, randomly chosen
network configuration of $N=1000$ neurons for $p=0.05$ and
$\tau_D=0.1$. In Fig.\ref{f:multatt}a presents each of the 471
different solutions that can be reached by exciting initially a single
neuron with all other neurons being at rest. As shown more clearly in
Fig.\ref{f:multatt}b, which gives the number of initial conditions
that lead to a solution with a given period (irrespective of their
firing rate and its standard deviation), only few values for the
period arise (note the logarithmic scale). In contrast, the standard
deviation of the firing rate (Fig.\ref{f:multatt}c) can take on many
different values, each representing a different temporal evolution of
the firing during the period. 

\begin{figure}
\begin{center}
\includegraphics[scale=0.4]{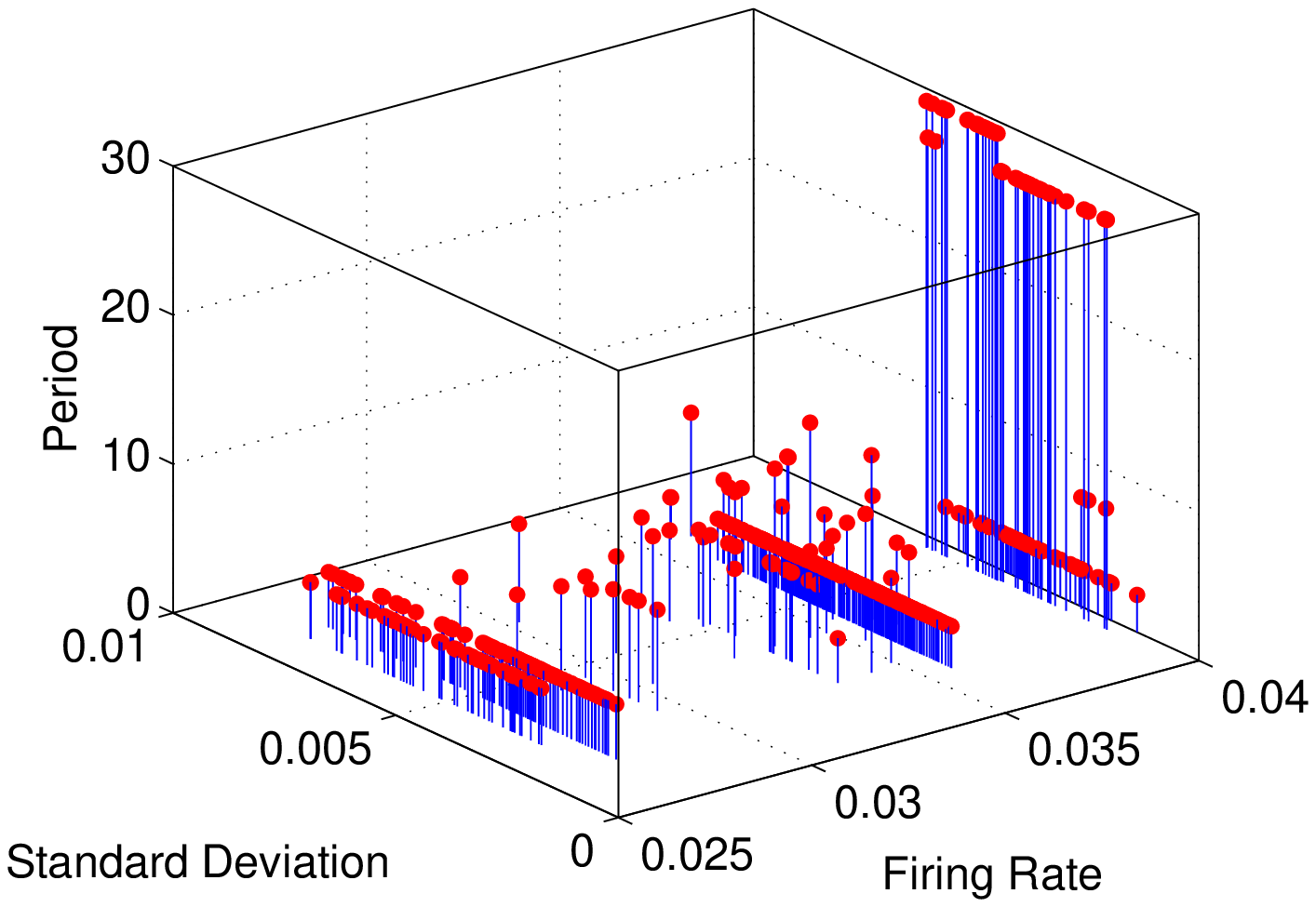}\\
\vspace*{.2cm}
\includegraphics[scale=0.4]{period_dist_nx1000_p0.05.eps}\\
\vspace*{.2cm}
\includegraphics[scale=0.4]{std_dist_nx1000_p0.05.eps}
\end{center}
\caption{Large number of attractors. a) Mean firing rate, 
its standard deviation, and period of all
471 attractors obtained for a single configuration for $N=1000$,
$p=0.05$, and $\tau_D=0.1$. b) Number of initial conditions leading to
an attractor with a given period. Many attractors, although distinct, 
have the same period. (note logarithmic scale). c) Number
of initial conditions leading to an attractor with a given standard
deviation of the firing rate.} 
\LB{f:multatt}
\end{figure}

The large number of attractors is typical for these networks as seen
in Fig.\ref{f:multatt-nx}, which presents the number of attractors as
a function of the network size averaged over 20 different
configurations (black circles). A large fraction of the attractors has
a quite small basin of attraction: within the restricted set of
initial conditions in which only a single neuron is excited, they are
reached from only a single initial condition (red squres). The overall
number of attractors seems to increase roughly linearly with system
size for larger values of $N$, while the number of attractors with
different periods (green diamonds) seems to grow more rapidly than
that. Since the duration of the transients grows with system size the
computation time grows faster than $N^2$. This precludes us from going to
significantly larger system sizes than shown in Fig.\ref{f:multatt-nx}
\footnote{For $N=4000$ the computation takes over 2 weeks on a desktop
PC.}, which would be necessary to get a reliable estimate for this
scaling. Preliminary computations indicate that allowing more general
initial conditions significantly increases the number of attractors
beyond those shown in Fig.\ref{f:multatt}. Thus, while the restricted
initial conditions employed in Fig.\ref{f:multatt-nx} suggest an
almost linear increase in the number of attractors with system size,
the full number of attractors may grow substantially faster. At this
point the origin for the exceedingly large number of different
attractors is not clear. In all-to-all coupled oscillator systems
factorially large numbers of attractors have been found due to the
permutation symmetry of that global coupling \cite{WiHa89}. The
small-world networks investigated here do not possess any such
symmetries.

\begin{figure} 
\begin{center}
\includegraphics[scale=0.4]{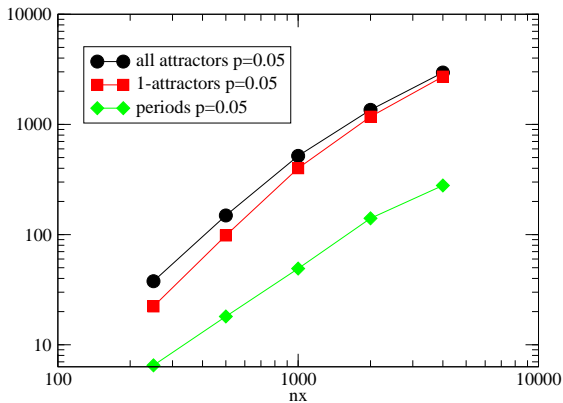}
\end{center}
\caption{Dependence on the system size. All attractors (circles),
attractors with basin of attraction 1 (squares), attractors with
different period (diamonds).
}  \LB{f:multatt-nx}
\end{figure}

As the value of $p$ increases, the distance traveled by the waves
before encountering a long-range connection decreases. Consequently,
the  waves of excitation spread throughout the network more rapidly.
This effect can be seen clearly in the progression of spatio-temporal
patterns in Fig\ref{f:increasep}(a-e). For sufficiently large values
of $p$ the activity spreads too quickly to sustain the activity, see
Fig.\ref{f:increasep}f. Since the long-range connections are made
randomly, the dynamics vary across network configurations. Thus, while
the overall likelihood of persistent activity should decrease with
increasing $p$, the actual network dynamics depend on the particular
network configuration.

The dynamical mechanism leading to the extinction of network activity
is  easily elucidated \cite{RoRi04}. Once a neuron has emitted a
spike, its voltage is reset, here to $0$.  As the voltage of the
neuron recovers to its resting potential, here equal to $V_\infty$,
the activity spreads through the network, eventually finding its way
back.  Once this occurs, the neuron receives synaptic input equal to
$g_{syn}$.  This input will be only sufficient to trigger a spike in
this neuron if the neuron has sufficiently recovered.  It is therefore
clear that the more rapidly the activity spreads, the less likely the
network will be to exhibit persistence. Whether or not this mechanism
leads to the extinction of activity in a given network depends on the
particular network topology. For fixed $p$ many configurations of the 
network can be simulated to calculate the percentage of network
configurations for which the  activity is extinguished, as shown in
the left inset of Fig.\ref{f:failures}.  In agreement with our
intuitive  argument, the likelihood of the activity failing for a
particular configuration drawn at random increases with increasing
$p$.  In fact, for large enough systems it seems that there is a
transition or, more precisely, a cross-over from a regime in which all
network configurations exhibit persistent activity, to one in which
the activity will always fail. The transition moves to larger values
of $p$ as the system size $N$ increases.

As discussed in \cite{RoRi04}, this transition can be captured in a
mean-field  approximation in which the return time is identical for
all neurons.  In this case the dependence on the long-range component
of the topology can be expressed as a function of $p$  alone.  Setting
the maximum return time $T_{A}$ (the time needed for the activity to
traverse the  entire network) equal to the recovery time yields an
upper bound for the density of  long-range connections at which a
transition from persistent activity to failure occurs 

\begin{equation}
T_{A}(p_{cr})=T_{R}^{(1)}. \LB{crit}
\end{equation}

An approximate form for $T_{A}$ is easily derived.  We assume that a
single neuron fires at time  $t=0$.  Given a density $p$ of shortcuts,
we expect a long-range connection will be reached after $1/p$ neurons
have fired, which occurs after a time $\tau_{D}/2p$ since there are
two wave fronts.  Four wave fronts are now present and so $2/p$
neurons will fire in a time $\tau_{D}/2p$ at which point on average
two new wave fronts are generated, and  so on.  During the $k^{th}$
cycle, the number of neurons excited is $2^{k-1}/p$.  After $n$ cycles
all the neurons have been  excited, which results in

\begin{equation}
\sum_{k=0}^{n-1}2^{k}=pN
\end{equation}
and leads to a time
\begin{equation}
T_{A}(p)\equiv n\frac{\tau_D}{2p}=\tau_{D}\frac{\ln{(1+pN)}}{2p\ln 2}. \LB{e:Ta}
\end{equation}

Equation (\ref{e:Ta}) is a purely geometric result, since the time it
takes for activity to traverse the entire extent of the network is
related trivially to the largest distance in the network. The
geometric mean-field properties of small-world networks have been
worked out by Newmann, Moore and Watts.  In \cite{NeMo00} they
calculate the  fraction of a small-world network covered by starting
at a single point and extending outwards a distance $r$ in both
directions (equivalent to waves having spread for a time $r\tau_{D}$
in our case) in a continuum limit.  In \cite{NeMo00} they refine the
continuum limit calculation  by taking into account two effects that
were omitted from their earlier calculation \cite{NeWa99}  and
from ours. Firstly, as they trace out the network, a jump via a
long-range connection may reach a part of the network that has already
been traced over and should therefore not be counted.  We interpret
this as activity spreading via long-range  connections, which is
injected into a neuron that has already fired but is not ready yet.  
Such a connection would therefore be ineffectual.  Secondly, when two
`traced-out' sections meet, they stop and no longer contribute.  We
interpret this as wavefronts which meet and annihilate and thus,
thereafter, do not contribute. Considering these two additional
mechanisms, they develop a two-component model that describes,
effectively, the fraction of the network covered and the number of
fronts.  The result, interpreted within the context of equation
(\ref{e:IF}), yields,
\begin{equation}
\sqrt{\Big(1+\frac{4}{pN}\Big)}\tanh{\Big[\sqrt{\Big(1+\frac{4}{pN}\Big)}\frac{pT_{A}(p)}
{2\tau_{D}}\Big]}=1.  \LB{Ta3}
\end{equation}
Setting $T_{A}(p_{cr})=T_{R}^{(1)}$, yields the density
$p_{cr}^{(MFT)}$ at the failure transition.  

The main panel in Fig.\ref{f:failures} shows the failure rates,
rescaled by the critical density $p_{cr}^{(MFT)}$ calculated from
Eq.(\ref{Ta3}). All failure curves intersect at the same value of $p$,
which defines therefore the transition point $p_{cr}$. Consistent with
the upper bound obtained in the mean field theory
$p_{cr}<p_{cr}^{(MFT)}$. Unfortunately, we need to point out that the
good quantitative agreement reported in \cite{RoRi04} was spurious;
there we incorrectly used $T_R$ instead of $T_R^{(1)}$ as the recovery
time in (\ref{Ta3}) (cf. Fig.\ref{f:failrefract}). For comparison with
the approximate mean field calculation the right inset shows the
failure curves rescaled by $p_{cr}$ as obtained from (\ref{e:Ta}). 

\begin{figure}
\begin{center}
\includegraphics[scale=0.3]{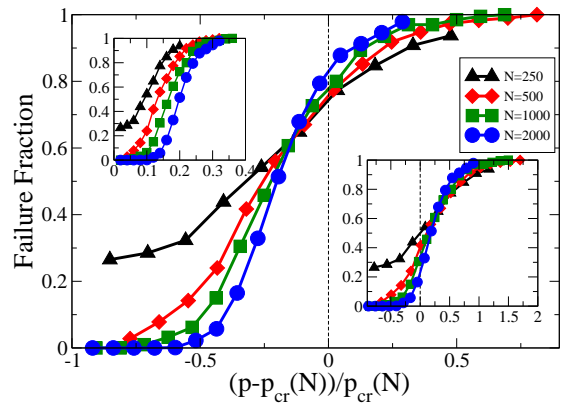}
\end{center}
\caption[]{Failure Transition as a function of system size $N$. 
Fraction of failing configurations rescaled by $p_{cr}$ from Eq.\ref{Ta3}.  
Left inset: unscaled data. Right inset: data rescaled by $p_{cr}$ from
Eq.\ref{e:Ta}.} \LB{f:failures}
\end{figure}

The central quantity determining the persistence or failure for a
given short-cut configuration is the recovery time $T_R^{(1)}$. It is
worth pointing out that this time is not the same as an absolute
refractory period $T_r$ of the neuron. While  $T_r$ is a property of
the neuron independent of its neighbors, $T_R^{(1)}$ depends strongly
on the strength of the coupling between the neurons. As long as the
refractory period is shorter than the recovery time it has little
effect on the persistence of the network. This is illustrated in
Fig.\ref{f:failrefract}, which shows the fraction of persisting
configurations for a network of neurons with an absolute refractory
period $T_r$ during which they do not respond to any input from other
neurons. During that time their membrane voltage $V$, however, still
relaxes towards its resting value. For $T_r<2\tau_D$ the second input
characteristic of propagating waves (cf. (\ref{e:Tr1})) is unaffected
and the relevant recovery time is $T_R^{(1)}$. The resulting failure
curve ($T_r=0$) corresponds to those in Fig.\ref{f:failures}. For
$2\tau_D<T_r$ the second input is suppressed. In this low-$p$ regime
the persistent states do not depend on neurons receiving any further
additional inputs before their firing (cf. Sec.\ref{s:disorder}) and
the failure transition is independent of $T_r$ as long as $2\tau_D<T_r
<T_R$. Only for $T_r>T_R$ does the transition depend on the absolute
refractory period. 

\begin{figure}
\begin{center}
\includegraphics[scale=0.5]{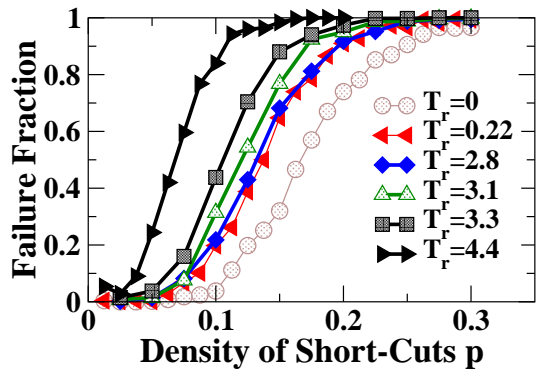}
\end{center}
\caption[]{Dependence of failure on refractory period $T_r$ for
$N=1000$, $\tau_D=0.1$ ($T_R=2.83$), and 500 configurations. }
\LB{f:failrefract}
\end{figure}

\section{The disordered regime: chaotic transients for slow waves}
\LB{s:disorder}

For small values of $p$ we have seen that the likelihood of persistent
activity is high.  In this regime, the spatio-temporal activity,
despite the complex, heterogeneous topology of the network itself, is
highly regular and most often periodic.  As $p$ increases a transition
in the likelihood  of persistent activity occurs and more and more
network configurations exhibit activity which peaks and then shuts
down. Interestingly, for large enough $\tau_{D}$ there is an
additional change in the behavior of the network as  $p$ increases,
see Fig.\ref{f:failuress}.  Specifically, one finds that while the
likelihood of failure initially increases for low $p$, as  described
in the previous section, it then turns downward again for larger
values of $p$, see the curves for $\tau_{D}=0.16$ and $0.18$ in
Fig.\ref{f:failuress}.  To understand why this is so, we first
describe the  spatio-temporal dynamics underlying this seemingly
persistent activity.

\begin{figure}
\begin{center}
\includegraphics[scale=0.5]{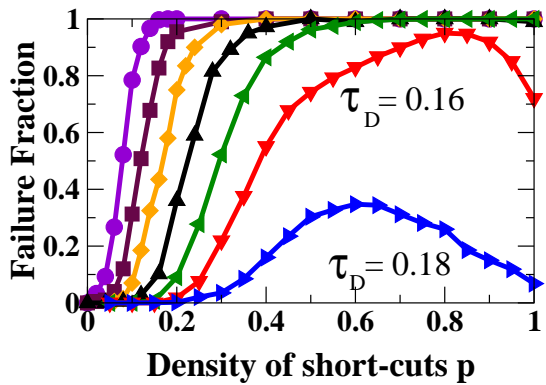}
\end{center}
\caption[]{Failure curves for different values of the delay
$\tau_{D}=0.06,0.08,0.1,0.12,0.14,0.16$.  Each symbol is the  average
of 2000 configurations.  For large enough delay, $\tau_{D}=0.16$ and
$0.18$, the likelihood of failure before a finite time $T^{*}=100$
becomes non-monotonic in $p$.} \LB{f:failuress}
\end{figure}

Fig.\ref{f:increasep2} shows four typical raster plots of the activity
seen for $\tau_{D}=0.16$ as a function of $p$.  For values of  $p$
below or near the theoretical transition to failure (panels (a)
through (c)), the activity is similar to that seen in
Fig.\ref{f:increasep}. In the regime beyond the maximum of the failure
curve (panel (d)) the activity is chaotic (cf.
Fig.\ref{f:oscidisorder}a below) and exhibits irregular population
spikes reflecting near-synchronous activity involving a large fraction of
the neurons.

\begin{figure}
\begin{center}
\includegraphics[scale=0.6]{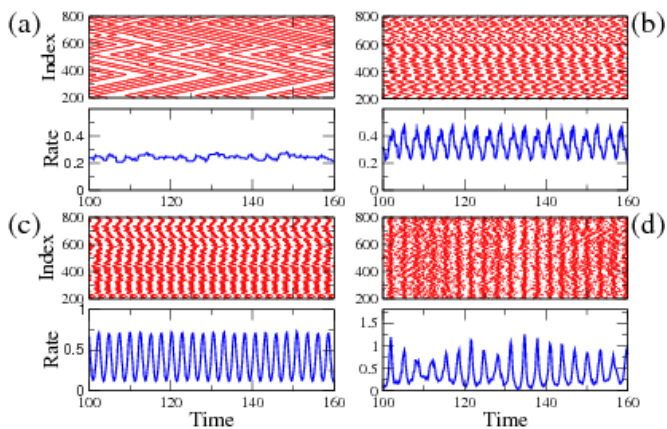}
\end{center}
\caption[]{Raster plots and firing rate for four typical
configurations for 'slow' waves, $\tau_{D}=0.16$.  
$p=0.01,0.2,0.4$ and $1.0$ for (a) through (d)
respectively.  The activity in the 'high' $p$ regime seen in (d) is
unlike that seen in the other three.  It is noisy and exhibits
synchronized population spikes.} \LB{f:increasep2}
\end{figure}

A more detailed, quantitative analysis for $\tau_D=0.18$ shows that
the change in behavior occurs already before the maximum of the
failure curve. For the fast waves (small $\tau_D$) the oscillation
amplitude as measured by the standard deviation of the firing rate
increases monotonically with $p$ (cf. Fig.\ref{f:osci}). For slower
waves it is, however, non-monotonic and decreases over the range $0.4
\le p \le 0.6$ (Fig.\ref{f:entropy} $\tau_D=0.18$). More instructive
yet is the dependence of the spectral entropy of the firing rate, 

\begin{equation} 
{\mathcal
E}=-\sum_\omega S(\omega)\,\ln S(\omega), 
\end{equation}

which measures the number of significant peaks in the power spectrum
${\mathcal S}(\omega)$. Its average over 200 configurations rises
significantly in this range of $p$, as shown in Fig. \ref{f:entropy},
indicating an increase in the complexity of the dynamics. The
variability of ${\cal E}$ across the configurations exhibits a (broad)
maximum in the transition region and reaches very small non-zero
values around ${\mathcal E}\sim 0.06$ in the strongly chaotic regime.
We have not investigated details of this evolution towards chaotic
dynamics, which depends on the individual configurations of the
short-cuts. 

\begin{figure}
\begin{center}
\includegraphics[scale=0.4]{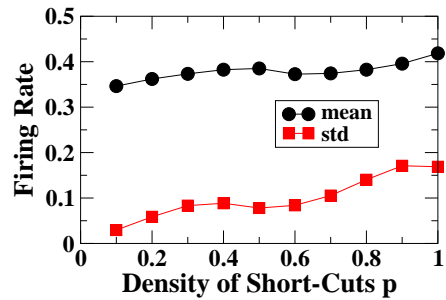}
\end{center}
\caption{
Oscillation amplitude depends non-monotonically on the
density of short-cuts. Mean and standard deviation of the firing rate
is shown for those configurations (out of 200) for which 
the activity persists for at least 15,000 steps. ($\tau_D=0.18$, $N=1000$).}
\LB{f:oscidisorder}
\end{figure}

\begin{figure}
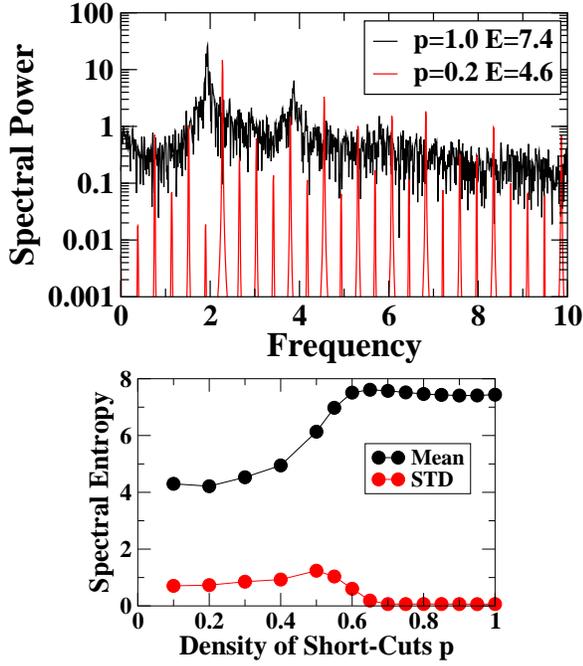

\begin{center}
\includegraphics[scale=0.5]{spectra_p1_p0.2.eps} 
\includegraphics[scale=0.4]{entropy_mean_sd_taud018.eps}
\end{center}
\caption{Temporal behavior becomes increasingly complex with
increasing $p$. a) Representative power spectra for $p=0.2$ 
(${\mathcal E}=4.6$) and $p=1.0$ (${\mathcal E}=7.4$). 
b) Spectral entropy $S$ (mean and standard deviation
across 200 configurations).  ($\tau_D=0.18$, $N=1000$ and 
15,000 time steps)}
\LB{f:entropy}
\end{figure}

What causes the dynamics to re-emerge as $p$ increases?  The answer
lies in the interplay between network topology and the delay
$\tau_{D}$.  In order to develop the mean-field model, eq.(\ref{crit})
we assumed that the firing of each neuron is limited by the recovery
time $T_R^{(1)}$. However, even for $p<1$ the construction of the
network allows for some neurons to receive more than one incoming
short-cut. For our network it can be shown \cite{Ro03} that for $p$
small, the fraction of neurons with two incoming short-cuts is
$s_{2}\sim p^{2}/2$, which is indeed negligible\footnote{For
simplicity we assume that the number of neurons with more than 2
inputs can be ignored and we compute the most likely rather than the
expected value}.  Thus, the mean-field results should hold if the
transition occurs at sufficiently low values of $p$.  However, for
$|p-1| \ll 1$ the fraction of neurons with two inputs becomes $s_{2}=
1-\sqrt{2}/2+(p-1)/2$ which is nearly $0.3$ for $p=1$.  Thus, a
significant  subpopulation of neurons likely receives several inputs
per cycle.  Such neurons would not be constrained by the recovery time
$T_{R}^{(1)}$ but rather would be primed to fire earlier, potentially
allowing the activity to persist where it otherwise  would fail.  We
can easily calculate the recovery time for a neuron $i$ which, by
virtue of several incoming connections, has received $n$ inputs since
its last firing

\begin{equation}
T_{R}^{(n)}(t_1,...,t_n)=\ln{\Bigg(\frac{V_\infty-g_{syn}\sum_{r=1}^{n}e^{t_{r}}}
{V_\infty+g_{syn}-1}\Bigg)}. \LB{e:Trni}
\end{equation}

Note that Eq.(\ref{e:Trni}) reduces to Eq.(\ref{e:Tr}) for $n=0$ and
to Eq.\ref{e:Tr1} for $n=1$ with $t_1=2\tau_D$. In general,
$T_{R}^{(n)}$ depends on the $n$ firing times of those neurons
providing input to neuron $i$, which are unknown. However, since
integrate-and-fire neurons become increasingly sensitive to their
input as time passes after their firing, the value
of $T_{R}(n)$ in Eq.(\ref{e:Trni}) is  bounded below by
$T_{R,min}^{(n)}$, which occurs when all $n$ inputs coincide at
$T_{R,min}^{(n)}$ itself. This time is given by

\begin{equation}
T_{R,min}^{(n)}(t_i=T_{R,min}^{(n)})=\ln{\Bigg(\frac{V_\infty}{V_\infty+ng_{syn}-1}\Bigg)},
\LB{trn}
\end{equation}

For small enough $p$ the inter-spike intervals (ISI) of almost all the
neurons are bounded below by $T_{R}^{(1)}$, which allows us to
calculate the time for activity to spread throughout the whole network
using a geometrical approach.  For higher values of $p$ there may be a
subset of neurons with shorter allowable ISI.  However, many neurons
will still only receive a single input per cycle and their activity
should reflect this fact.  This is borne out in the distribution
function for the ISI shown in Fig.\ref{f:bridge}. `Fast' spikes with
${\rm ISI}<T_r^{(1)}$ occur appreciably only for $p\ge 0.6$ and
increase further in frequency with increasing short-cut density. Our
previous analysis \cite{RoRi04} showed that the spikes with ${\rm
ISI}>T_R^{(1)}$ occur in population bursts with no such spikes in
between. In the absence of other spikes the activity would die out
during these periods. However, due to the large number of short-cuts
there is a substantial number of neurons that have received multiple
inputs and are consequently primed to carry over the activity to the
next cycle, thereby allowing the `slow' spiking neurons to recover.

\begin{figure}
\begin{center}
\includegraphics[scale=0.4]{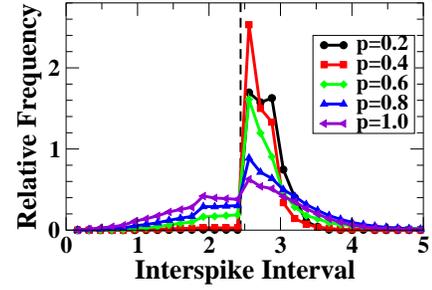}
\end{center}
\caption[]{Distribution of interspike intervals (ISI) for $\tau_{D}=0.16$ 
and a range of $p$. For low $p$ all ISI are above the recovery period
$T^{(1)}_R$ (dashed line). For $p\ge0.6$ increasingly more multiple 
inputs allow shorter ISI.}
\LB{f:bridge}
\end{figure}

Increasing the delay $\tau_D$ contributes in a number of ways towards
bridging low-activity periods. For larger $\tau_D$ the failure
transition is shifted towards larger short-cut densities enhancing the
number of multiple-input neurons significantly. At the same time, in
order to bridge the time between the relatively short return time
$T_A$ and the recovery time $T_R^{(1)}$ of the slow neurons fewer fast
neurons with multiple input are needed if the delay is longer.
Moreover, for the integrate-and-fire neurons (\ref{e:IF}) later inputs
have a  stronger impact on the recovery period than earlier ones (cf.
Eq.(\ref{e:Trni})). With increasing $\tau_D$ all inputs are shifted to
later times relative to the last spike of the respective neuron, which
reduces the recovery time significantly. The increased delay is,
however, not necessary to reach the regime of prolonged activity. As
shown in Fig.\ref{f:largeN} for fixed delay $\tau_D=1.4$, increasing
the system size from $N=1,000$ to $N=16,000$ shifts the failure
transition to sufficiently large $p$ that the number of neurons with
multiple inputs is sufficient to bridge the gap in slow-neuron
activity even for this lower delay time. 

\begin{figure}
\begin{center}
\includegraphics[scale=0.45]{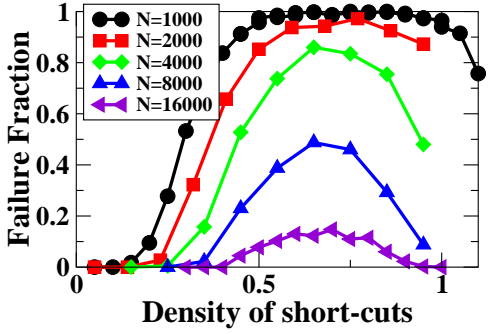}
\end{center}
\caption{Prolonged activity in large systems. Failure rates up to a
final time $T^*=28$ for $\tau_D=0.14$ (400 configurations).  }
\LB{f:largeN}
\end{figure}

For low values of $p$ the spatio-temporal dynamics are most often
periodic. In those cases the dynamics can truly be called persistent.
For $p\sim 1$, the chaotic nature of the dynamics precludes such a
clear assessment and, in fact, failure is possible even after very
long times. Since persistent activity relies on a bridging of the
quiescent period by a string of neurons that have multiple incoming
short-cuts it is necessary that these short-cuts are actually
activated during the previous cycle and at suitable times. Thus, while
in one cycle the activity during the burst may have been able to
excite such a chain the different activity pattern in the next cycle
may fail to do so and the activity could die out. Indeed, we find that
for these large short-cut densities the activity eventually fails for
essentially all configurations. Examples of such long-lived transients
are shown in Fig.\ref{f:changetaud}, where firing rates are shown for
the same configuration of short-cuts but increasing values of
$\tau_D$. Clearly, the delay has a strong influence on the duration of
the transient (note the change in scale on the $x$-axis).

\begin{figure}
\begin{center}
\includegraphics[scale=0.3]{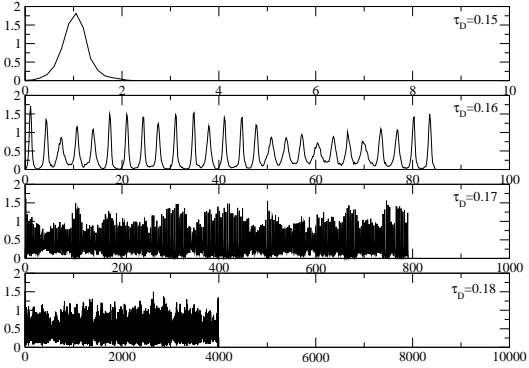}
\end{center}
\caption[]{The firing rate for four values of the delay $\tau_{D}$,
with $p=1$.  The network architecture and initial condition are
identical for all four simulations, only the delay has been changed.
Failure tends to occur at longer times for larger delays.}
\LB{f:changetaud}
\end{figure}

While overall there is a trend for the lifetimes of the transient
activity to increase with increasing delay, the actual dependence on
$\tau_D$ is more subtle. Fig.\ref{f:failuresdisttauD} shows, as a
function of $\tau_{D}$, the fraction of configurations for which the
activity fails before the final time $T^*$ is reached. The dependence
on $\tau_D$ exhibits an amazing degree of structure. Most surprising
is the finding that an increase in $\tau_D$ does not always decrease
the number of failures, but can in fact enhance the probability of
failure. These changes can occur over very small intervals in $\tau_D$
as shown more clearly in Fig.\ref{f:failuresdisttauD}b. This fine
structure is reminiscent of `resonances' or `windows'  in which the
delay is $\tau_D$ `optimal'.  While details of the mechanism
underlying this structure are not understood yet, it is clear that the
dependence on $\tau_D$ reflects the significance of the ratio
$\tau_D/T_R$. This is further illustrated in
Fig.\ref{f:failuresdisttauD2} where the failure rates are shown for a
reduced number of neurons, $N=500$. While the failure rates are higher
overall in the smaller system,  the locations of the windows in
$\tau_D$ are not substantially altered. If however, the recovery time
is reduced from $T_R=2.83$ to $T_R=2.79$ by increasing the synaptic
strength from $g=0.2$ to $g=0.202$, the windows are clearly shifted to
lower values of $\tau_D$.

\begin{figure}
\begin{center}
\includegraphics[scale=0.5]{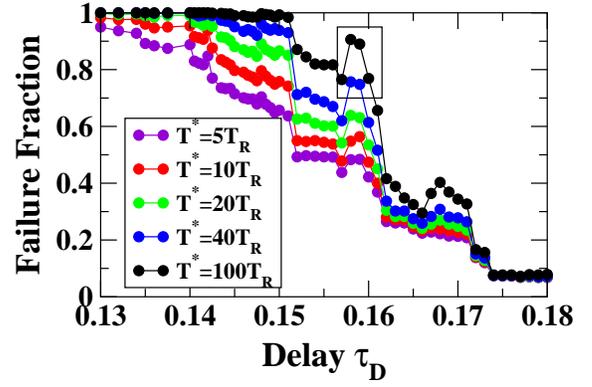}

\vspace{.2cm}
\includegraphics[scale=0.5]{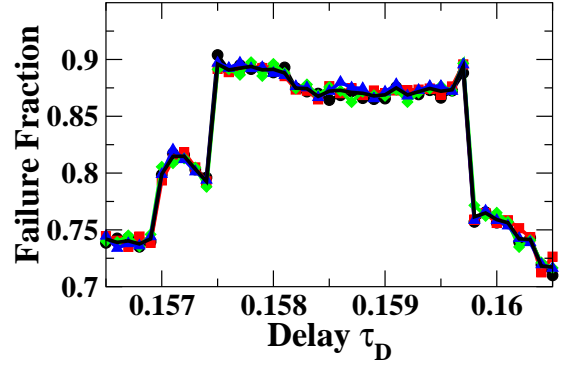}
\end{center}
\caption[]{a) Failure rates for a range of final times $T^*=5 T_R$,
$10T_R$, $20T_R$, $40T_R$, $100T_R$ based on 2000
configurations with $p=1$ and $N=1,000$. The range covered in b) is marked by a box. 
b) Fine structure in the failure rate for $T^*=100T_R$
based on four runs with 8000 configurations each. Black line gives
average over the 4 runs.}
\LB{f:failuresdisttauD}
\end{figure}

\begin{figure}
\begin{center}
\includegraphics[scale=0.5]{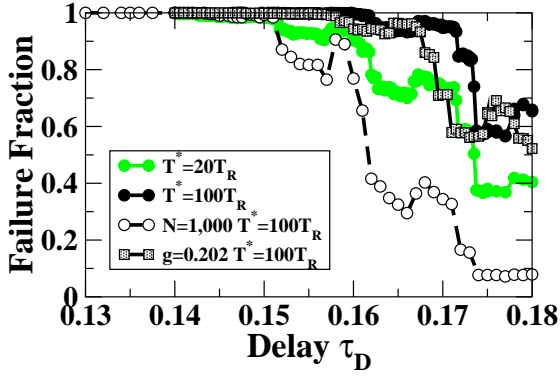}
\end{center}
\caption[]{Failure rates based on 8000
configurations with $p=1$ for $N=500$ for $g=0.2$ and $g=0.202$. For
comparison also the case $N=1000$ is shown (cf. Fig.\ref{f:failuresdisttauD}).}
\LB{f:failuresdisttauD2}
\end{figure}

To assess whether any significant fraction of the configurations leads
to truly persistent activity we consider the failure fraction ${\cal
F}$ as a function of the final time $T^*$, with the aim to extrapolate
to $T^*\rightarrow \infty$.  Fig.\ref{f:failuresdisttime} shows the
fraction of persistent configurations, $1- {\cal F}(T^*)$, for a range
of delays $\tau_D$ on a logarithmic scale. As may have been
aniticipated from the non-monotonicity seen in
Fig.\ref{f:failuresdisttauD}, the fraction of failing configurations
is largest for $\tau_D=0.167$ over the whole range of times monitored
and a very rapid drop in the failure rate is seen when going from
$\tau_D=0.17$ to $\tau_D=0.18$. Considering the long-term behavior it
is apparent from Fig.\ref{f:failuresdisttauD} that the decay is
non-exponential. 

\begin{figure}
\begin{center}
\includegraphics[scale=0.7]{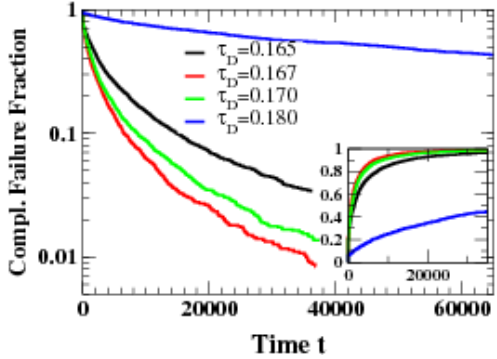}
\end{center}
\caption[]{Complementary failure fraction $1-{\cal F}$ as a function of 
time for a range of delays $\tau_D$. The decay is non-exponential.
Inset: Failure fraction ${\mathcal F}$ for the same data.}
\LB{f:failuresdisttime}
\end{figure}

To obtain an approximate, analytic form for the failure fraction
${\cal F}(T^*)$ we consider first - for a fixed value of $p$ - a
single fixed configuration. The duration of the activity before
failure will then depend on the specific initial condition chosen.  In
these simulations we choose 2,000 random initial conditions $V=V_0+\xi$  with
$V_0=0.85$ and $-0.5<\xi<0.5$ picked from a uniform distribution. The
resulting distribution for the failure times exhibits exponential
behavior for large times. This allows the extraction of a
characteristic failure time $T$ associated with this configuration. To
reduce the computational effort these simulations are done in a
smaller system with $N=200$. The exponential distribution suggests
that the chaotic dynamics effectively lead to a fixed probability for
the activity to die out after each population spike.

Across all configurations this leads then to a distribution
$\rho (T)$ of characteristic failure times, in terms of which - averaged
over many configurations - the fraction of failures up to a time
$T^*$  is given by

\begin{equation}
{\mathcal F}(T^*)=\int_{0}^{\infty}dT \rho (T)(1-\beta
e^{-\frac{T^*}{T}}). \LB{cumfail}
\end{equation}

Extracting the characteristic lifetimes $T$ of 50,000 configurations
we find that the distribution $\rho (T)$ can also be well fit by an
exponential function for long lifetimes,

\begin{equation} 
\lim_{T\to\infty}\rho (T)\propto e^{-\alpha T}. 
\LB{rho} 
\end{equation}

\begin{figure}
\begin{center}
\includegraphics[scale=0.5]{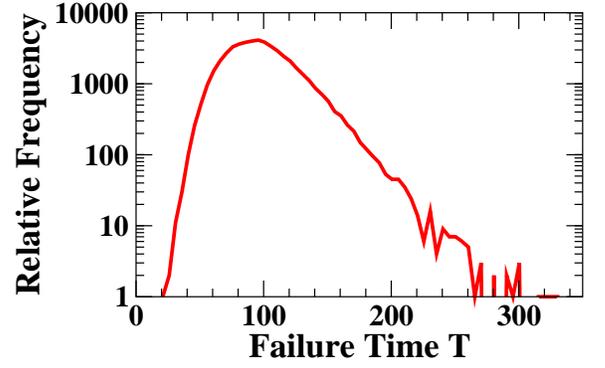}
\end{center}
\caption[]{The distribution of characteristic failure times across
50,000 configurations. The distribution decays exponentially for 
long lifetimes. $N=200$, $p=1$, $\tau_D=0.16$}
\LB{f:lifetimedist}
\end{figure}

Fig.\ref{f:lifetimedist} shows the result for $\tau_D=0.16$ and $p=1$
($N=200$). Inserting the asymptotic behavior of Eq.(\ref{rho}) into
Eq.(\ref{cumfail}) yields for large $t$

\begin{equation}
{\cal F}(t)=1-2\beta\sqrt{\alpha t}K_{1}(2\sqrt{\alpha t}) \LB{e:bessel}
\end{equation}

where $K_{1}(x)$ is the modified Bessel function of the second kind of
first order.  For long times Eq.\ref{e:bessel} can be expressed as

\begin{equation} 
{\mathcal F}(t)\sim 1-\beta\sqrt{\pi}(\alpha t)^{1/4}e^{-2(\alpha t)^{1/2}}
\end{equation} 

displaying stretched exponential behavior. As Fig.\ref{f:besselfit}
shows, Eq.\ref{e:bessel} provides quite a good fit to the time
dependence of the failure fraction. For $\tau_D=0.18$ the fit
corresponding to $\alpha=0.67\cdot 10^{-5}$ and $\beta=0.91$ is good
over essentially the whole range of times. For $\tau_D=0.165$ (inset)
the fit is not quite as good; its curvature seems smaller than that of
the data. Starting at $t=3,400$ one obtains $\alpha=0.12\cdot10^{-3}$
and $\beta=0.64$. This fit overshoots the data for large times (blue
dotted line in Fig.\ref{f:besselfit}). The fit with $\alpha=0.10\cdot
10^{-3}$ and $\beta=0.54$ (red dashed line) reduces the overshoot but is not
as good for smaller times. 

\begin{figure} 
\begin{center}
\includegraphics[scale=0.5]{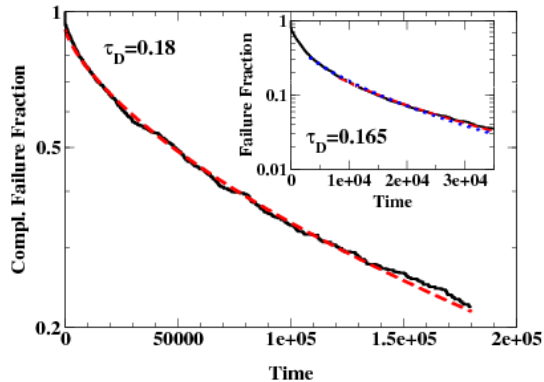} 
\end{center} 
\caption[]{Complementary failure fraction $1-{\cal F}$ for $p=1$ and 
$\tau_D=0.18$ (inset for $\tau_D=0.165$). The dashed red and dotted blue 
lines show fits to Eq.(\ref{e:bessel}).} \LB{f:besselfit} 
\end{figure}

\section{Conclusion}
\LB{s:concl}

In this paper we have used a minimal model to study the influence of
the network topology on the dynamics of excitable elements. The
network consisted of local connections between neighboring elements on
a ring that are supplemented  by randomly chosen short-cuts connecting
distant elements. Having neural systems in mind we have taken the
short-cuts to provide uni-directional connections rather than the
bi-directional ones that would arise in epidemic contexts
\cite{KuAb01} or in regular diffusive processes. 

Depending on the density of short-cuts and the speed of the waves
propagating through this network we find three regimes. For low, but
non-zero density of short-cuts the activity persists for essentially
all network configurations when starting from an initial excitation of
a single neuron. The activity is predominantly periodic and the mean
firing rate of these states shows only little dependence on the wave
speed or the density of short-cuts once $p\ge 0.05$ and it is
reasonably close to the maximal firing rate allowed by the recovery
period of the neurons. The recovery period is, however, not to be
confused with an absolute refractory period; rather, it is the time
after which synaptic input of the strength used in the computations is
sufficient to trigger a new spike. This recovery period can be much
longer than the usual absolute refractory period, for instance, if the
neurons exhibit slow after-hyperpolarization as it underlies the slow
oscillations ($< 1$Hz) observed in vivo in cat \cite{AmSt95} and in
cortical slices of ferret \cite{SaMc00}. There the relative refractory
period induced by the after-hyperpolarization can last as long as a
few seconds. The dependence of the propagation speed of such slow
oscillations on the connectivity has been studied in cortical models
\cite{CoSa03,SaCo05}. No true short-cuts were employed, instead the
spatial width of the Gaussian giving the probability that two neurons
are connected was varied \cite{SaCo05} or a tri-modal probability
distribution was used to capture a certain patchy connectivity in
cortex \cite{CoSa03}. As expected, the speed was found to increase
with the width and it was conjectured that this connectivity
dependence is the origin for the large difference observed for the
slow waves in olfactory cortex and neocortex, respectively
\cite{SaCo05}.

Even though for small numbers of short-cuts the long-time dynamics 
are periodic, the over-all behavior of the system can be quite complex
due to the large number of stably coexisting solutions for a given
network configuration. Whether the number of attractors grows as fast
with the system size as in globally coupled oscillator networks, where
the permutation symmetry essentially leads to a factorial growth of
the number of attractors and to attractor crowding \cite{WiHa89}, is
not known at this point. In fact, so far we have not been able to
reach saturation of the number of attractors with increasing number of
different initial conditions and the mechanism underlying this large
number of attractors is not apparent yet. It is clear, however, that
noise will induce a  persistent switching between these different
attractors \cite{Ro03}. 

As the density is increased the number of network configurations that
allow such a persistent activity decreases until essentially no
persistent activity is possible anymore. For fast waves the transients
after a localized excitation consist of a single population burst. For
slower waves, however, for which the cross-over to complete failure
occurs at yet larger short-cut densities, the transients can be
exceedingly long comprising thousands of population bursts. The times
at which the failures in activity averaged over different
configurations occur are distributed according to a stretched
exponential. It arises from the exponential distribution of the
lifetimes characterizing each configuration. The mechanism responsible
for these long transients presumably differs from that operating in
diluted random networks of pulse-coupled oscillators \cite{ZuTi04}.

While the fraction of failing configurations up to a given time
follows a decreasing overall trend with decreasing wave speed, it
exhibits intricate fine structure that includes even sharp, 
resonance-like increases of the failure fraction with decreasing wave
speeds. This is surprising, since naively one might expect that a
decrease in wave speed would allow additional, shorter loops to
contribute to the activity and therefore enhance the chances for
persistence. However, the result indicates that the activation of one
loop can, in fact, make other, previously active loops impossible and
thus induce failure. While it is clear that such a switching between
loops can occur with increasing $\tau_D$, this mechanism is not yet
understood in any detail. 

The cross-over to failure, which can be understood analytically in
detail based on a mean-field theory for the size of these idealized
small-world networks, provides the basis for understanding a number of
recent investigations employing more complex neural network models
\cite{NeCl04,ShTs06}. 

In \cite{NeCl04} the connection between network connectivity and
epilepsy in hippocampus was investigated by considering noisy and more
elaborate versions of our model. It was found that for low numbers of
short-cuts the noise drives only low-level activity, which was
associated with normal behavior. With increasing short-cut density the
activity strongly increased due to the recruitment of many more
neurons by a noise-driven event. This activity was likened to seizing
activity. This regime corresponds to the connectivities supporting
persistent activity. Yet further increases in the short-cut density
were found to induce bursting dynamics in which irregular bursts that
involved a large fraction of all neurons are separated by periods of
quiet. One would expect such a behavior for connectivities in the
failing regime in which each noise-triggered event leads to a single
population spike that brings essentially all neurons into the recovery
period. 

In \cite{ShTs06} the system was driven by a set of localized
pace-maker neurons. It was found that the short-cuts can lead over a
number of driving cycles to the build-up of bursts during which a
large fraction of neurons fire within a small time window. The time to
build up such bursts and the time between them was found to decrease
with short-cut density. Again, the appearance of such bursts is
related to the failing configurations in systems without driving. As
expected from this observation the bursting behavior was supplanted by
persistent activity when the wave speed was reduced (cf.
Fig.\ref{f:failuress}). The slow build-up towards the burst, however,
is relatively specific to the Morris-Lecar model employed in that
study for the individual neurons \cite{ShTs06}.

We gratefully acknowledge support by NSF through grant DMS-0309657
and the IGERT program "Dynamics of Complex Systems in Science and
Engineering" (DGE-9987577) (HR,AR,SM) and by the EU under grant
MRTN-CT-2004-005728 (SM). 


\end{document}